\documentclass[runningheads]{llncs}
\usepackage{graphicx}
\usepackage[T1]{fontenc}
\usepackage{verbatimbox,lipsum}
\usepackage{subcaption}
\usepackage[utf8]{inputenc}
\usepackage{listings}
\usepackage{pifont}
\begin{document}
%


\title{Scholarly Knowledge Extraction \\from Published Software Packages}

%
%
\author{Muhammad Haris\inst{1}\orcidID{0000-0002-5071-1658} \and
Markus Stocker\inst{2,1}\orcidID{0000-0001-5492-3212} \and
Sören Auer\inst{2,1}\orcidID{0000-0002-0698-2864}
}

\authorrunning{Haris et al.}

\institute{L3S Research Center, Leibniz University Hannover 30167, Hannover, Germany\\
 \email {haris@l3s.de} \and
TIB---Leibniz Information Centre for Science and Technology, Germany
\email{\{markus.stocker,auer\}@tib.eu}}

\maketitle              

\begin{abstract} 
A plethora of scientific software packages are published in repositories, e.g., Zenodo and figshare. These software packages are crucial for the reproducibility of published research. As an additional route to scholarly knowledge graph construction, we propose an approach for automated extraction of machine actionable (structured) scholarly knowledge from published software packages by static analysis of their (meta)data and contents (in particular scripts in languages such as Python). The approach can be summarized as follows. First, we extract metadata information (software description, programming languages, related references) from software packages by leveraging the Software Metadata Extraction Framework (SOMEF) and the GitHub API. Second, we analyze the extracted metadata to find the research articles associated with the corresponding software repository. Third, for software contained in published packages, we create and analyze the Abstract Syntax Tree (AST) representation to extract information about the procedures performed on data. Fourth, we search the extracted information in the full text of related articles to constrain the extracted information to scholarly knowledge, i.e. information published in the scholarly literature. Finally, we publish the extracted machine actionable scholarly knowledge in the Open Research Knowledge Graph (ORKG).

\end{abstract}
\section{Introduction}
\label{s:introduction}
A variety of general and domain-specific knowledge graphs have been proposed to represent (scholarly) knowledge in a structured manner~\cite{spacecraftknowledgegraph,Zhao2018Architecture}. General purpose knowledge graphs include DBpedia\footnote{\url{https://www.dbpedia.org}}~\cite{dbpedia}, Wikidata\footnote{\url{https://www.wikidata.org/wiki/Wikidata:Main\_Page}}~\cite{wikidata}, YAGO~\cite{yago}, etc., whereas domain-specific infrastructures include approaches in Cultural Heritage~\cite{domainspecific}, KnowLife in Life Sciences~\cite{knowlife}, Hi-Knowledge in Invasion Biology\footnote{\url{https://hi-knowledge.org}}~\cite{heger2013conceptual,enders2020conceptual}, COVID-19 Air Quality Data Collection\footnote{\url{https://covid-aqs.fz-juelich.de}}, Papers With Code in Machine Learning\footnote{\url{https://paperswithcode.org}}, Cooperation Databank in Social Sciences\footnote{\url{https://cooperationdatabank.org}}~\cite{spadaro2020cooperation}, among others. In addition, knowledge graph technologies have also been employed to describe software packages in a structured manner~\cite{kelly,Abdelaziz_toolkit}.

Extending the state-of-the-art, we propose an approach for scholarly knowledge extraction from published software packages by static analysis of package contents, i.e., (meta-)data and software (in particular, Python scripts), and represent the extracted knowledge in a knowledge graph. The main purpose of this knowledge graph is to capture information about the materials and methods used in scholarly work described in research articles.

We address the following research question: Can structured scholarly knowledge be automatically extracted from published software packages? Our approach consists of the following steps:
\begin{enumerate}
    \item \textit{Mining software packages} deposited in Zenodo\footnote{\url{https://zenodo.org}} using its REST API\footnote{\url{https://developers.zenodo.org}} and analyzing the API response to extract the linked metadata information, i.e, associated scholarly articles. We complement the approach by leveraging the Software Metadata Extraction Framework (SOMEF) to parse the README files and extract other related metadata information (i.e., software name, description, used programming languages).
    \item \textit{Perform static code analysis} to extract information about the procedures performed on data. We utilize Abstract Syntax Tree (AST) representations to statically analyze program code and identify operations performed on data.
    \item \textit{Identify scholarly knowledge} by performing keyword-based search of extracted information in article full text. Thus, among all the information extracted from software packages we identify that which is scholarly knowledge.
    \item \textit{Construct a knowledge graph} of scholarly knowledge extracted from software packages. For this purpose, we leverage the Open Research Knowledge Graph (ORKG)\footnote{\url{https://www.orkg.org/orkg/}}~\cite{orkg}, a production research infrastructure that supports producing and publishing machine actionable scholarly knowledge.
\end{enumerate}

\section{Related Work}
\label{s:related-work}
Several approaches have been suggested to retrieve metadata from software repositories. Mao et al.~\cite{mao} proposed the Software Metadata Extraction Framework (SOMEF) to extract metadata from software packages published on GitHub. Specifically, the framework employs machine learning-based methods to extract repository name, software description, citations, reference URLs, etc. from README files and to represent the metadata in structured formats (JSON-LD, JSON and RDF). SOMEF was later extended to extract additional metadata and auxiliary files (e.g., Notebooks, Dockerfiles) from software packages~\cite{kelly}. Moreover, the extended work also supports creating a knowledge graph of parsed metadata, thus improving search of software deposited in repositories. Abdelaziz et al.~\cite{Abdelaziz2020ADO} proposed CodeBreaker, a knowledge graph with information about 1.3 million Python scripts published on GitHub. The graph was embedded in an IDE to recommend code functions while writing software. Similarly, GraphGen4Code~\cite{Abdelaziz_toolkit} is a knowledge graph with information about software included in GitHub repositories. It was generated by analyzing the functionalities of Python scripts and linking them with the natural language artefacts (documentation and forum discussions on StackOverflow and StackExchange). The knowledge graph contains 2 billion triples.
Several other machine learning-based approaches for searching~\cite{husain2020codesearchnet} software scripts and summarization~\cite{ahmad-etal-2020-transformer,iyer-etal-2016-summarizing} have been proposed. The Pydriller~\cite{PyDriller} and GitPython\footnote{\url{https://github.com/gitpython-developers/GitPython}} frameworks were proposed to mine information from GitHub repositories, including source code, commits, branch differences, etc. Similarly, ModelMine~\cite{modelmine} mines and analyzes models included in repositories. Vagavolu et al.~\cite{Vagavolu} presented an approach that leverages Code2vec~\cite{le2014distributed} and includes semantic graphs with Abstract Syntax Tree (AST) for performing different software engineering tasks.~\cite{allamanis2017learning} presented an AST based-approach for code representation and considered code data flow mechanisms to suggest code improvements.

\section{Methodology}
\label{s:methodology}
\begin{figure*}[t!]
  \includegraphics[width=\textwidth]{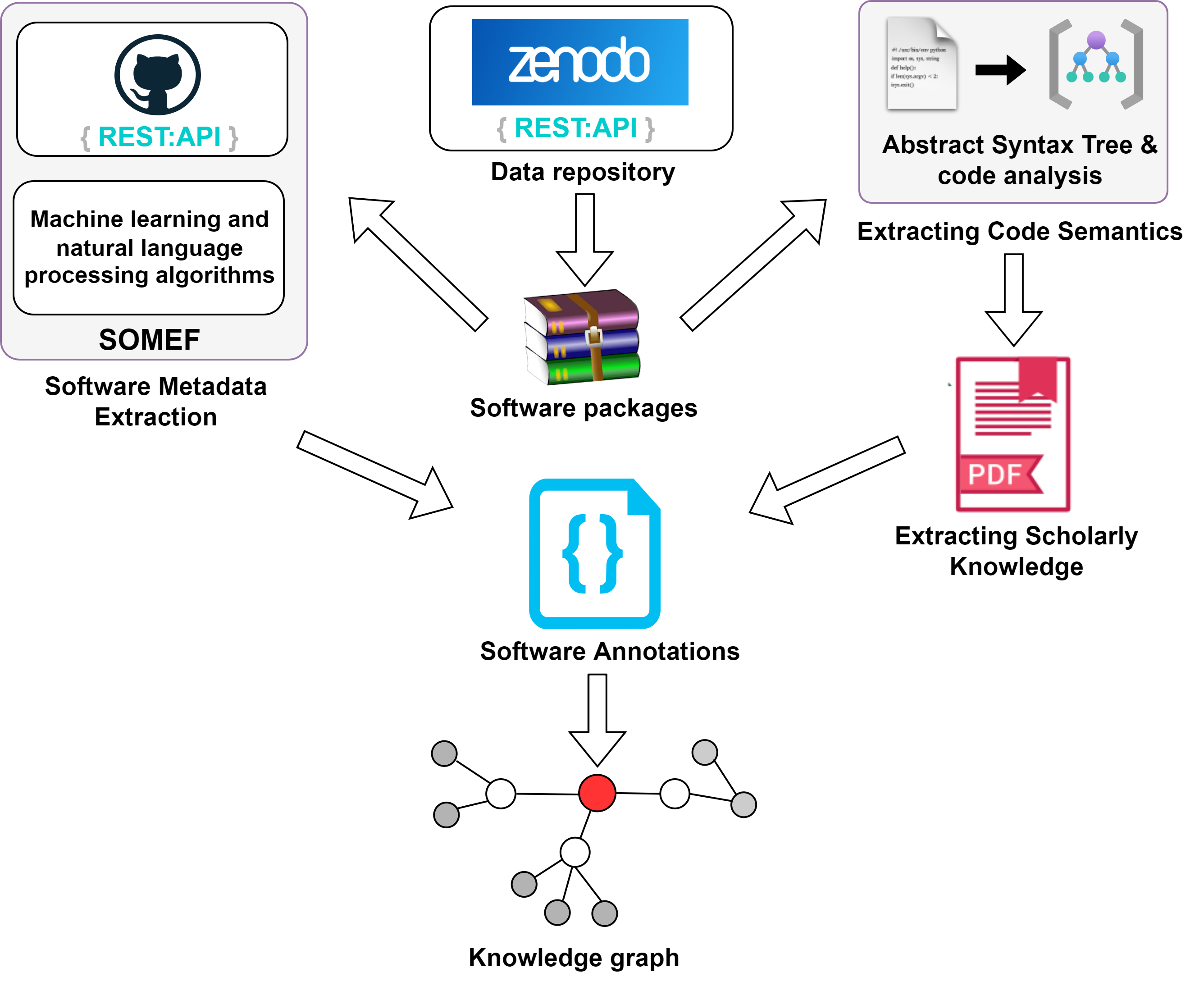}
  \caption{Pipeline for constructing a knowledge graph of scholarly knowledge extracted from software packages: 1) Mining software packages from the Zenodo repository using its REST API; 2) Extracting software metadata by analyzing the Zenodo API results as well as the GitHub API, using SOMEF; 3) Performing static code analysis using AST representations of software to extract code semantics, in particular operations on data; 4) Performing keywords-based search in article full texts to identify scholarly knowledge; 5) Knowledge graph construction with scholarly knowledge extracted from software packages.}
  \label{fig1}
\end{figure*}

In this section, we present our methodology for automatically extracting \emph{scholarly} knowledge from software packages and building a knowledge graph from the extracted meta(data). Figure~\ref{fig1} provides an overview of the key components. 

\subsection{Mining Software Packages}
We mine software packages from the Zenodo repository by leveraging its REST API. The metadata of each package is analyzed to retrieve its DOI and metadata about related versions and associated scholarly articles. The versions of software packages are retrieved by interpreting \texttt{relation: isVersionOf} metadata, whereas the DOI of the linked article, if available, is fetched using the \texttt{relation: cites} or \texttt{relation: isSupplementTo} metadata. We also leverage the Software Metadata Extraction Framework (SOMEF) and GitHub API to extract additional metadata from software packages, in particular software name, description, used programming languages, GitHub URL. Since not all software packages include the \texttt{cites} or \texttt{isSupplementTo} relations in metadata, we utilize SOMEF to parse the README files of software packages as an additional approach to extract the DOI of the related scholarly article.

\begin{figure*}[t!]
  \includegraphics[width=\textwidth]{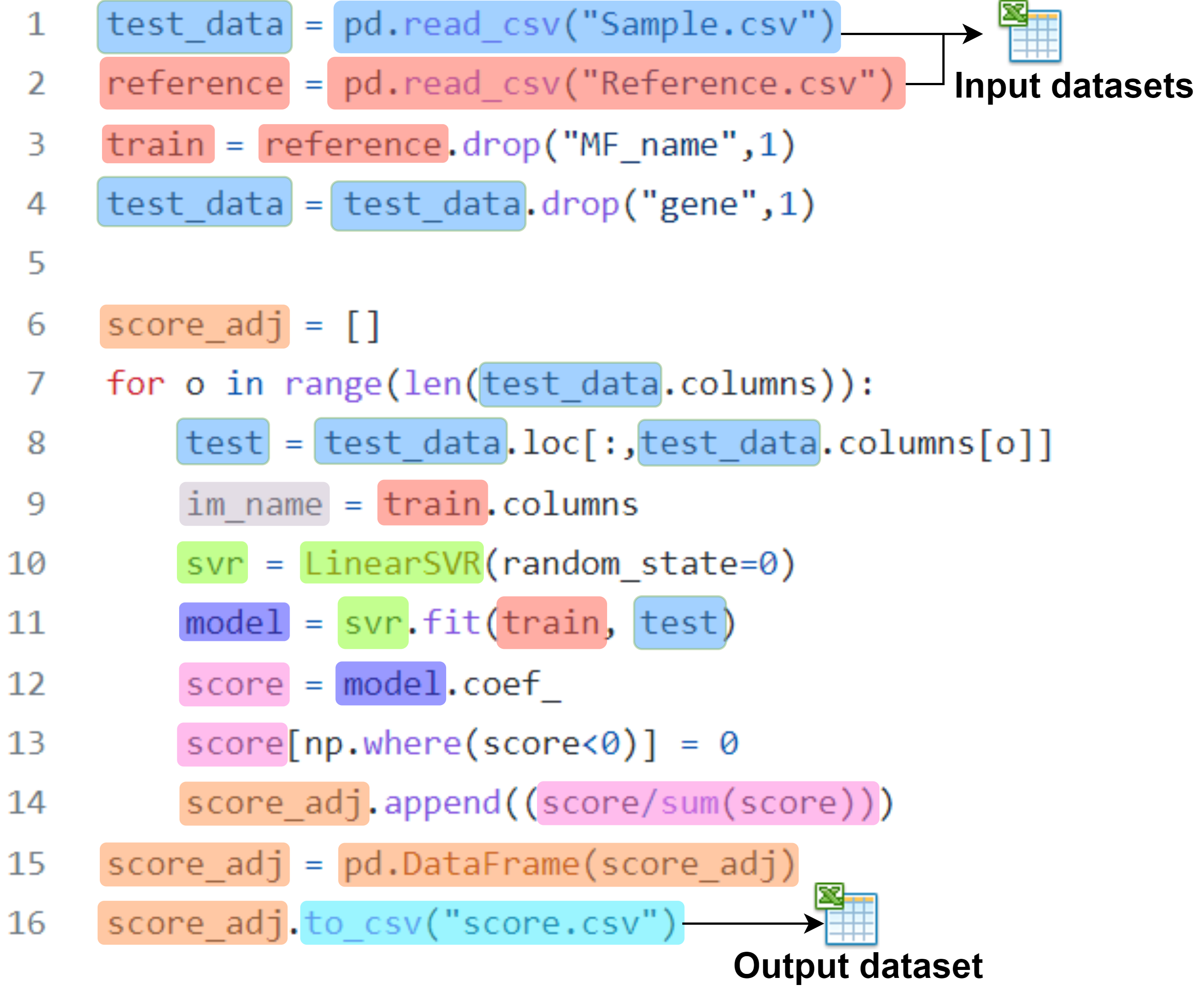}
  \caption{Static code analysis: Exemplary Python script (shortened) included in a software package. The script lines highlighted with same color show different procedural changes that a particular variable has undergone.}
  \label{fig3}
\end{figure*}

\subsubsection{Static Code Analysis}
We utilize Abstract Syntax Tree (AST) representations for static analysis of Python scripts included in software packages. AST provides structured representations of scripts, omitting unnecessary syntactic details (e.g., semicolons, commas, and comments). Our goal is to extract information about the data used in scripts and the procedures performed on that data. Our developed Python-based module sequentially reads the scripts contained in software packages and generates the AST. The implemented procedures and variables are tokenized and represented as nodes in the tree, which facilitates the analysis of the code flow. Thus, by traversing the tree we extracts the information about the data used in the scripts, the procedures performed on the data and, if available, the output data.
\begin{figure*}[t!]
  \includegraphics[width=\textwidth]{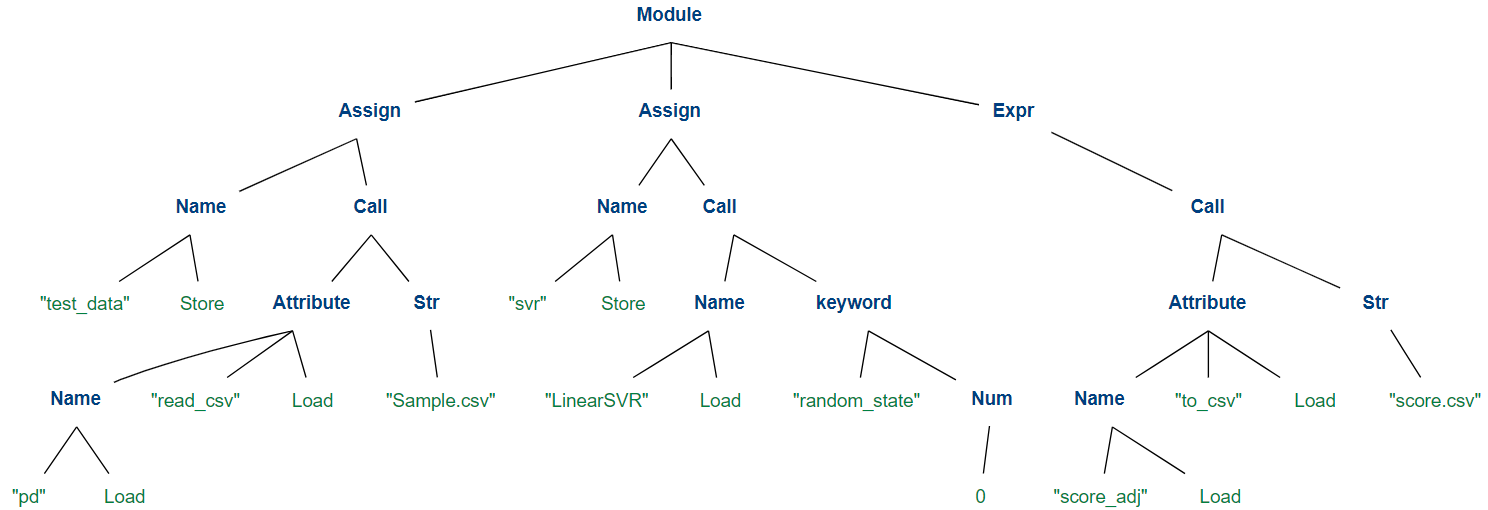}
  \caption{Abstract Syntax Tree (AST) of the script shown in Fig.~\ref{fig3}. For simplicity, the AST is shown only for Lines 1, 10 and 16.}
  \label{fig6}
\end{figure*}
Fig.~\ref{fig3} shows the Python script included in the software package\footnote{\url{https://zenodo.org/record/5874955}}. The script shows an example in which \texttt{Sample.csv} and \texttt{Reference.csv} used as input data, then the operation \texttt{LinearSVR} is performed on the data, and finally the resulting data \texttt{score.csv} is generated.

Fig.~\ref{fig6} shows the AST of the Python script (Fig.~\ref{fig3}) created using a suitable Python library\footnote{\url{https://docs.python.org/3/library/ast.html}}. For simplicity, we show the AST of lines 1, 10, and 16. In the tree structure, the name of the node represents the functionality of each line of the script. For example, line 1 performs a task that reads data and \texttt{assigns} it to a variable. Therefore, the relevant node in the tree is labelled \texttt{Assign}. We retrieve all leaf nodes since they represent variables, their values, and procedures. Analyzing these script semantics, we can then find the flow of data between procedures. We investigate the flow of variables that contain the input data, i.e., examining which operations used a particular variable as a parameter. 

\subsection{Identifying Scholarly Knowledge}
Not all information extracted from software packages and AST-analyzed program code is scholarly knowledge. Information is scholarly knowledge if it is included in a scholarly article. Hence, we filter the information extracted from software packages for information referred to in the article citing the software package. For this, we employ keyword-based search. Specifically, we search for the terms extracted in AST-analyzed program code in the related article full text. Assuming that the DOI of the related article has been identified, we fetch the PDF version of the article by utilizing the Unpaywall REST API\footnote{\url{https://api.unpaywall.org/v2/10.1186/s12920-019-0613-5?email=unpaywall\_01@example.com}}. We make use of the Unpaywall API because, contrary to DOI metadata, it provides the URL to the PDF version of scholarly articles. In our example (Fig.~\ref{fig3}), the extracted terms (\texttt{Sample}, \texttt{Reference}, \texttt{read\_csv}, \texttt{LinearSVR}, \texttt{svr.fit}, and \texttt{to\_csv}) are searched in the PDF and we find \texttt{Sample}, \texttt{Reference} and \texttt{LinearSVR} are cited in the scholarly article. We thus assume that the extracted information is scholarly knowledge.

\subsection{Knowledge Graph Construction}
We now construct the knowledge graph with the scholarly knowledge obtained in the analysis of software packages. For this, we leverage the Open Research Knowledge Graph (ORKG)~\cite{orkg}. The ORKG aims to represent scholarly articles in a machine actionable and structured form. Abstractly speaking, the ORKG represents research contributions describing key results, the materials and methods used to obtain the results, and the addressed research problem.

\begin{figure*}[t!]
  \includegraphics[width=\textwidth]{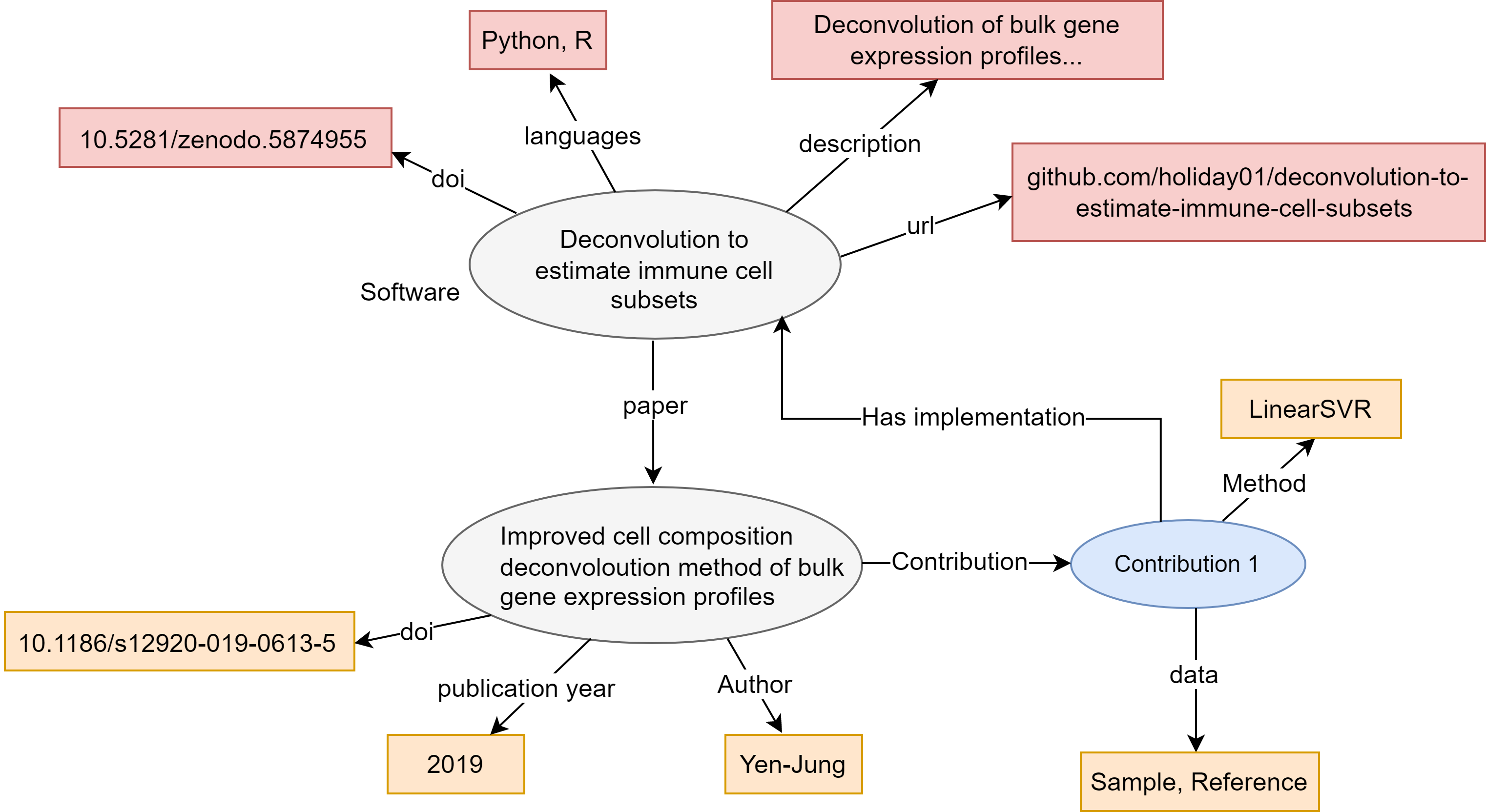}
  \caption{Knowledge graph depicting the scholarly knowledge extracted from a software package related to an article, describing key aspects (e.g., method used) of a research contribution of the work described in the article.}
  \label{fig4}
\end{figure*}
The scholarly information extracted from software packages in organized in triples and ingesting into ORKG using its REST API. Fig.~\ref{fig4} shows the resulting knowledge graph for a paper and its research contribution\footnote{\url{https://orkg.org/paper/R209873}}. The figure also shows the metadata of corresponding software package\footnote{\url{https://orkg.org/content-type/Software/R209880}}.
\section{Results and Discussion}
\label{s:discussion}
At the time of writing, there are more than 80,000 software packages available on Zenodo. To expedite the execution process, we discard packages larger than 400 MB. We thus consider 52,236 software packages. We further process only those software packages that are also available on GitHub, that is 40,239 packages. We analyze the metadata of the software packages and the respective README files and find a total of 6221 research articles, of which 642 articles are associated with the related software packages in metadata through the \texttt{cites} or \texttt{isSupplementTo} relations. The remaining 5579 articles are extracted by analyzing the README files of the software packages using SOMEF. We only analyze software packages that include Python scripts and have linked scholarly articles, that is 2172 packages. Table~\ref{table1} summarizes the statistics.

\begin{table*}[]
\caption{Statistics about the (scholarly) information extracted from software packages.}
\centering
\begin{tabular}{|p{5cm}|l|l|l|l|l|l|l|l|}
\hline
  \textbf{Entity} & \textbf{Total} \\ \hline
\textit{Software package} & 52236    \\ \hline
\textit{Paper} & \multicolumn{1}{p{5cm}|}{Explicit links in metadata: 642; SOMEF-based link extraction: 5579 (Total: 6221)}  \\ \hline
\textit{GitHub URL} & 40,239   \\ \hline
\textit{Python-based software packages, linked with articles} & 2172 \\ \hline
\textit{Analyzed Python scripts} & 67,936   \\ \hline

\end{tabular}
\label{table1}
\end{table*}

Out of 6221 articles, 4328 are described in ORKG because for the remaining articles the DOIs in README files are not parsed correctly. The articles added to ORKG include ORKG research contribution descriptions linking the software package and including information about computational methods and data used in research extracted by analyzing the software packages.

\paragraph{Software semantics and Named Entity Recognition (NER) models} There exist numerous approaches for the extraction of scholarly knowledge from articles using machine learning and natural language processing, including scientific named entity recognition~\cite{Jiang,Coreference} and sentence classification~\cite{Crossdomain}. These approaches process the entire text to extract the essential entities in scholarly articles, which is costly in terms of data collection and training. Moreover, the approaches require large training data to achieve acceptable performance. We argue that extracting scholarly knowledge from software packages as proposed here is a significant step towards automated and cheap construction of scholarly knowledge graphs. Instead of extracting scholarly entities from full texts using machine learning models, the scholarly knowledge is extracted from related software packages with more structured data.

\paragraph{Future directions} In future work, we aim to develop a pipeline that will automatically execute the software packages that contain scholarly knowledge. Such an approach can be integrated into software repositories (zenodo, figshare) to automatically execute the published software and determine whether the (extracted) scholarly knowledge is reproducible.

\section{Conclusions}
\label{s:conclusion}
Our work is an important step towards automated and scalable mining of scholarly knowledge from published software packages and creating the knowledge graph using the extracted data. The resulting knowledge graph holds the links between articles and software packages, as well as and most interestingly descriptions of the computational methods and materials used in research work presented in articles. Evaluated on zenodo, our approach can be extended to other repositories, e.g., figshare, as well as software in languages other than Python, e.g., R, Java, Javascript, and C++---potentially further increasing the number of articles and related scholarly knowledge added to ORKG.

\section*{Acknowledgment}
This work was co-funded by the European Research Council for the project ScienceGRAPH (Grant agreement ID: 819536) and TIB--Leibniz Information Centre for Science and Technology.

\bibliographystyle{splncs04}
\bibliography{paper}
\end{document}